\documentclass[12pt]{article}

\begin{document}

\input epsf

\begin{center}
{\bf \Large Radio Detection of High Energy Particles: Coherence
Versus Multiple Scales}
\end{center}

\vspace{1cm}

\begin{center}
{\large Roman V. Buniy and John P. Ralston}
\end{center}

\begin{center}
{\large \it Department of
Physics and Astronomy}\\{\large \it University of Kansas, Lawrence, KS
66045}
\end{center}

\vspace{0.5cm}

{\bf Abstract:} Radio Cherenkov emission underlies detection of high
energy particles via a signal growing like the particle energy-squared. 
Cosmic ray-induced electromagnetic showers are a primary application. 
While many studies have treated the phenomenon approximately, none have
attempted to incorporate all the physical scales involved in problems with
time- or spatially-evolving charges.  We find it is possible to decompose
the calculated fields into the product of a form factor, characterizing a
moving charge distribution, multiplying a general integral which depends
on the charge evolution.  In circumstances of interest for cosmic ray
physics, the resulting expressions can be evaluated explicitly in terms of
a few parameters obtainable from shower codes.  The classic issues of
Fraunhofer and Fresnel zones play a crucial role in the coherence. 

\section{Introduction}

Coherent radio Cherenkov emission is a remarkably effective method for
detecting high energy particles.  The history of the effect goes back to
Jelly, who first asked whether cosmic ray air showers might produce a
radio signal \cite{Allan}.  Askaryan \cite{Askaryan} subsequently
predicted a net charge imbalance in air showers, and coherent radio power
scaling like the energy of the shower squared.  Substantial radio emission
from atmospheric electromagnetic cascades was observed more than 30 years
ago.  \cite{Jelly, Allan}.  Progress in ultra-high energy air showers has
sparked renewed interest, and new observations of radio pulses have been
reported recently \cite{Rosner}.  The current pilot project RICE
\cite{fmr, RICE} uses radio Cherenkov emission to detect $100 TeV$ and
higher energy neutrinos in Antarctic ice.  The radio Cherenkov signal is
the most efficient known mechanism \cite{fmr,mkr, RICE, Provorov, Price}
for detecting
neutrinos of $100TeV$ and above in solid media, yielding detection volumes
of order $ 1km^{3}$ per radio detector for $PeV$ neutrinos on ice targets.
At $PeV$ energies and above, the neutrino interaction cross sections offer
fascinating new tests of Standard Model physics \cite {sigma} and new
physics \cite {gzknus}.  Tomography of the Earth is also possible with
$PeV$-scale neutrinos\cite {tomo}. Radio Cherenkov signals have also
been used to search recently for neutrinos and cosmic rays with energies
upwards of $10^{20}eV$ impinging on the Moon\cite{Moon}.

Cherenkov radiation is also an intrinsically interesting and beautiful
physical phenomenon.  Coherence is a basic feature of electrodynamics,
and the coherent enhancement of Cherenkov radiation in the microwave
region has been observed in the laboratory \cite{Takahashi, Wake}.
The Argonne wake-field acceleration project\cite{Wake} has
successfully generated extremely large microwave field-strengths by
manipulating coherent radiation from an intense electron beam.

Despite a long history, the previous literature apparently does not
contain a careful treatment of evolving charge distributions, such as
those of electromagnetic showers in air or ice, which incorporates all
important features of the problem.  The problem is intricate because of a
multitude of scales.  When an electromagnetic shower evolves, it produces
a pancake of charge with a finite thickness, a finite width, probed at a
finite wavelength of radiation, and for a finite distance over which the
shower is big.  All this occurs at a finite distance from the detector.
Results on evolving and finite-sized charge distributions are few.  Tamm
\cite{Tamm} grappled with the problem of a charged particle on a track of
limited length in the early days of the theory.  Askaryan \cite{Askaryan}
anticipated a coherence cut-off in air showers at high frequencies of
order the inverse pancake size, imposed somewhat by hand.
Allan\cite{Allan} gave physical arguments and order of magnitude estimates
based on one of Feynman's electrodynamic formulas.  Kahn and Lerche
\cite{KL} attempted to resolve the coherence issue using superpositions of
infinite tracks.  McKay and Ralston \cite{mkrlpm} and Alvarez-Mu\~{n}iz and
Zas ($AZ$)  \cite{AZ} considered the influence of the $LPM$ effect at
ultra-high energies.  Zas, Halsen, and Stanev ($ZHS$) \cite{ZHS}, and
$AZ$ reported results from summing asymptotic far-fields
track-by-track in Monte Carlo calculations of great complexity.

We present an approach which incorporates all the scales and allows a
general analysis.  Main results include an expression for the electric
field in a factorized form.  The ``factorization" occurs when distance
scales can be separated: the characteristic size of the moving charge
distribution must be substantially smaller than the scale over which
the charge develops.  This condition is well satisfied for all cosmic
ray applications we have examined.  A {\it form factor} characterizes
the moving charge distribution, which multiplies a charge-evolution
integral.  Not all the scales decouple: subtleties coming under the
classic description of {\it Fraunhofer} and {\it Fresnel} zones need
careful treatment.  Finally, the generic situation can be summarized
by analytic formulae.  This is indispensable given the large parameter
space.  For example, the numerous and varied numerical plots obtained
from immense Monte Carlos can be summarized by a few parameters.  With
the parameters fixed, predictions can be made for any number of
circumstances.

\section{Consequences of Coherence}

Before beginning analysis we review a few basics. The well-known
Frank-Tamm (1937) formula uses an exact solution to Maxwell's equations
for a uniformly moving charge on an infinitely long track. The solution is
kinematic and can be obtained by the trick of boosting the charge to a
speed faster than light in the medium. Extension to a track of finite
length has pitfalls.  Tamm's 1939 finite-track formula\cite{Tamm} assumes
a uniform charge $e$ traveling at a uniform velocity $v$ along the
$z$-axis for $-L/2< z < L/2$. Tamm gives the energy loss $dP$ per angular
frequency $d\omega$ per solid angle $d\Omega$:
\begin{equation}\frac{d^{2}\!P}{d\omega
d\Omega}=\frac{ne^{2}}{4\pi^{2}c^{3}}(\omega
L)^{2}\sin^{2}{\theta}\frac{\sin^{2}{X}}{X^2},\label{Tamm} \end{equation}
where $X=n\omega L/(2c)(\cos{\theta_{c}}-\cos{\theta})$ and
$\cos{\theta_{c}}=c/(nv)$; $n$ is the index of refraction.\footnote{We
diverge from current practice and use the special symbol $c$ to represent
the speed of light, otherwise known as ``1''.} This formula has been
cited in the high energy physics literature, and used to interpret
experiments observing Cherenkov radiation in the millimeter wavelength
range \cite{Takahashi}.

Tamm's finite-track formula includes two competing and distinct
physical processes: the Cherenkov radiation of a uniformly moving
charge, and bremmstrahlung or acceleration radiation from charges
modeled as starting and stopping instantly at the track's endpoints.
The interference of the sudden start and stop contributions with the
straight line contribution leads to strong oscillations in the angular
distribution.  Compared to a typical high energy process the
acceleration at the endpoints is fake, that is, the Tamm model is
unreliable.  This is because a charge created by pair production is
accompanied by an opposite charge, which coherently shields the pair
from radiating until the oppositely charged partners gradually
separate.  This ``Perkins effect" has been observed \cite{Perkins} and
is closely related to the coherence phenomena in QCD of color
transparency \cite{Color transparency}.  At the end of the
Cherenkov processes, charges also do not stop instantly, but
instead slow gradually to subluminal speeds.  While the slowing
has stochastic elements, it is better approximated by a uniform
decelleration than by a catastrophic disappearance of charge.
Of course, the evolution of a cosmic ray shower over many radiation lengths
and involving billions of particles is an even smoother
macroscopic process.  Thus the Tamm formula and
related approximations may misrepresent the physics if the
artificial treatment of the endpoints play a major role.
Conversely, experimental situations with conditions close to
those assumed by the Tamm formula can be constructed: Takahashi
{\it et al.} \cite{Takahashi} report on the sudden appearance of
a charge in a cavity with metallic boundary conditions, leading
to a strong mixing of Cherenkov and boundary-condition effects.

\subsection{Fraunhofer versus Fresnel}

Our study uncovers another, deeper problem with certain
asymptotic assumptions of the Tamm-type approach. In a typical
application of Cherenkov radiation in high energy physics, we
might have a track of length $L\sim1m$, observed at a distance
$R\sim1m$, and in the optical regime with $R/\lambda \sim
L/\lambda>10^6$. The application to radio detection in ice might
have $R\sim100-1000 m$, $\lambda\sim0.1-1m$, $L\sim 10m$, with
$L/\lambda\sim 10-100$ with $R/\lambda$ even greater. Cosmic ray
air showers develop and are observed over tens of kilometers. In
all cases, all lengths are large in units of the wavelength.
Given $L$ large enough for the acceleration contributions to be
small, the Tamm formula, Eq.~(\ref{Tamm}), might appear ideal at
first sight. Indeed for $\omega L\rightarrow\infty$, the
$\sin^{2}{X}/X^{2}$ distribution approaches $2\pi c/(n\omega
L)\delta(\cos{\theta}-\cos{\theta_{c}})$. Integrating over angles
we recover the well-tested Frank-Tamm result for an infinitely
long track, $$\frac{d^{2}P}{d\omega
dL}=\omega\frac{e^{2}}{c^{2}}\sin^{2}{\theta_{c}}.$$

Yet the Tamm formula is quite inapplicable to such problems.  This is
evident from the formula's prediction that the radiated energy will be
concentrated in coordinate space at $\theta=\theta_{c}$, up to a small
width due to diffraction.  For many of the physical situations cited, the
energy is actually spread rather uniformly over the length of a cylinder
surrounding the charge's trajectory.  This is a broad angular distribution
extending over angles $\Delta\theta\sim L/R$, where $R$ is the distance to the
receiver.  The Tamm formula, or any asymptotic far field approximation,
does not depend on the distance $R$, and cannot describe this simple truth.
True enough, the {\it momenta} (wave numbers) of photons have directions
that may be peaked at $\theta_k\sim\theta_{c}$, but this is not the same
thing as the {\it power density} $dP/d\Omega$ seen on a sphere surrounding
the system going like $\delta(\cos{\theta}-\cos{\theta_c})$.  There is no
paradox: if one fixes $L$ arbitrarily large, and then moves to an
asymptotically distant location $R\rightarrow\infty$, the photons traveling
at the Cherenkov angle will appear to come from a point source whose
angular size is diffraction-limited.  The Tamm formula is derived by taking
the limit $R\rightarrow\infty$; once taken, the case of finite $L/R$ is
unavailable.

To see this breakdown from a different perspective, one can use simple
dimensional analysis and geometrical reasoning.  The Fourier transform
of the electric field $E_ {\omega}$ has dimensions of mass.  Cherenkov
radiation for the ``long uniformly moving track" has cylindrical
symmetry.  At cylindrical radius $\rho$ from the track, the energy per
length $2\pi\rho E^{2}(\rho)$ remains constant.  Dimensional analysis
plus the cylindrical symmetry forces the electric field to go like
$\sqrt { \omega/\rho}$.  This is nothing like the usual radiation from
accelerated charges which has fields falling like $1/R$ in
three-dimensional space. And this peculiarity applies out to arbitrarily
large distance $\rho$, provided the track is long enough.  But if
$\rho$ is taken so large that the radiation appears to emerge from a
point source, the $E$ field must fall like $L \omega/\rho\sim L\omega
/R$.  (The factor of $L$ comes from the linear power per unit length
dependence.  Momentarily we will examine this in more detail.)

The breakdown of Tamm's formula is thus due to an interchange of
limits. Tamm's formula is obtained by making the {\it Fraunhofer
approximation}, which fails under a broad range of finite
track-lengths. The {\it Fresnel zone} describes a complementary
far-field region where the Fraunhofer approximation must be
modified. The basic physics of the Fresnel zone for Cherenkov
radiation is elementary but requires some care.

\subsection {The Coherence Zone}

Consider (Figure 1) a charge moving on a straight line. Let $R(t)$ be 
the instantaneous distance from the
charge to the observation point. Information propagating in the
medium at speed $c_m$ will arrive simultaneously from the track if 
$\partial R/\partial t=c_m$. This is the Cherenkov
condition: $\partial R/\partial t=v\cos{\theta}=c_m$ for velocity
$\vec v$ oriented at angle $\theta$ relative to the direction
$\hat R$. Note that $R(t)$ is the radius from the charge to the
point, not the vector position.

Due to the geometry of the track and observation point, uniform motion
produces acceleration of $R(t)$.  If $\partial R/\partial t$ were
constant, the fields arriving would all be in phase for the whole
track length.  However, the acceleration relative to the observation
point produces an extra radial change of order $\Delta R= 1/2
(\partial^2 R/\partial t^2 )(\Delta t)^{2}$.  Coherence of modes of
wavelength $\lambda$ is then maintained only over a finite region of
$\Delta R < \lambda.$ Since $\partial^2 R/\partial
t^2=v^2\sin^{2}{\theta}/ R$, we solve to find the condition $\Delta
t_{coh}<\sqrt{R\lambda}/(v\sin{\theta}) $.  Equivalently, there is a
finite spatial coherence region for given $R$, given wavenumber
$k=2\pi/\lambda$, namely $$\Delta z_{coh}<\sqrt{R/(k\sin^2{\theta})}
$$ over which the ``sonic boom" of radiation is built coherently.
Since $\Delta z_{coh}\sim\sqrt{R}$, the coherence zone grows to
infinite size as $R\rightarrow\infty$: but this limit cannot be taken
carelessly.

\subsection{Coherence of Evolving Charge Distributions}

We now return to the emission from an electromagnetic shower or other
time-evolving charge distribution. To a reasonable approximation, the
number of particles in a highly relativistic shower scales like the
primary energy divided
by a suitable low energy threshold. The charge imbalance near the
shower maximum is of order $20\%\, e$ of the total
number of particles. These numbers have been confirmed over and over,
with each generation of numerical simulation contributing further
detail. The origin of the emitted Cherenkov power going like the
shower-energy squared is basic electrodynamics: the electric
field will scale like the charge, and the radiated power scales like the
electric field-squared.

The evolving shower has a finite length scale $a$ over which it is
near its maximum, and radiating copiously.  This length scale, known
as the ``longitudinal spread'' in cosmic ray physics \cite {rossi}, is
akin to the length scale $L$ of Tamm's approach but represents a
smooth onset and decline of maximum power.  The shower maximum-length
scale $a$ is determined by the material, and is conceptually distinct
from the shower's total depth to reach the maximum (which goes like
the logarithm of the energy) or the charged pancake size (which is fairly
constant once the shower is developed.) A cartoon of these ideas is
given in Figure 1.

There are then two characteristic limits.  Suppose the longitudinal
spread of the shower is ``short" compared to the coherence length,
$a\ll\Delta z_{coh}$.  Then coherence is maintained over the whole
range where the current is appreciable.  The amplitude is proportional
to the total length $a$ over which the current was ``on'', times
$1/R$.  This is then the $R\rightarrow \infty$ limit, or Fraunhofer
approximation, and looks like normal radiation.  From dimensional
analysis (and here we recall the discussion earlier), $E_{\omega} \sim
a \omega/R$.

However, in the limit $a\ge\Delta z_{coh}$, the coherence length is
not as big as the longitudinal
spread, and coherence only exists
over the smaller of the two.  Adding amplitudes only over the region
$\Delta z_{coh}$ and weighted by $1/R$, the $E_{\omega }$ field
goes like $\omega\Delta z_{coh}/R=\sqrt{\omega/R}$.  This behavior is
rather different from the previous case: indeed Cherenkov radiation is
fundamentally a Fresnel-zone effect, as seen by the $1/\sqrt{R}$
dependence of the fields.

Both the Fresnel and Fraunhofer limits are far-field approximations,
in the sense that $kR\gg1$ is assumed.  The subtlety lies in the
dimensionless ratio $$\eta=(a/\Delta z_{coh})^{2}
=\frac{ka^{2}}{R}\sin^{2}{\theta}$$ which controls how the limit
$R\rightarrow\infty$ is taken. Confusion on this point is easy; one
has $ R/\lambda \sim10^{4}$ in the same regime, and yet $R$ is not
large enough for a ``large $R$" Fraunhofer approximation to apply,
exactly because the term ``large R" is undefined until the limit
parameter $\eta$ is specified.  In the RICE experiment one typically
has $a\sim 1-2 m$, $\omega\sim 100-1000 MHz$, and $R\sim10^{3}m$, so
$\eta<1$ holds.  Extension to closer observation points, or to
energies where the $LPM$ effect can give a much larger $a$, makes
$\eta\gg 1$ possible.  Partly due to the obscurity of the coherence
criteria, the Fraunhofer approximation has received much attention in
the previous literature \cite {ZHS,AZ}, except for those estimates
using fields with "cylindrical" symmetry \cite
{zm,mkr,jelley,Jackson}.

\subsection{General Set-Up}

Let $\vec {E}_{\omega}(x)$ be the time-Fourier transform of the
$E$ field, with similar notation for other fields.  The Maxwell
equations for a dielectric medium are\\ $\vec{\nabla}\cdot
\vec{D}_{\omega}=4\pi\rho, c\vec{\nabla}\times\vec{B}_{\omega}
=4\pi\vec{J}_{\omega}-i\omega\vec{D}_{\omega},
\vec{\nabla}\cdot\vec{B}_{\omega}=0,
c\vec{\nabla}\times\vec{E}_{\omega}=i\omega\vec{B}_{\omega},$ where
$\vec{D}_{\omega}(x)=\epsilon(\omega)\vec{E}_{\omega}(x)$.  There is a
wave equation for the vector potential $A^{\mu}_{\omega}(x)$, given by
$c(\nabla^{2}+k^{2})A^{\mu}_{\omega}(x) =-4\pi J^{\mu}_{\omega}(x)$,
with $k=\omega\sqrt{\epsilon}/c$.  Then we have
\begin{equation}c\vec{A}_{\omega}(\vec{x})
=\int\!\!d^{3}\!x'\,\frac{\exp{(ik|\vec{x}
-\vec{x}\,'|)}}{|\vec{x}-\vec{x}\,'|}\int\!\!dt'\,\exp{(i\omega
t')}\vec{J}(t',\vec{x}\,').\label{AA}\end{equation} The
$4$-potential $A^{\mu}=(A^{0},\vec{A})$ has been defined in a
generalized Lorentz gauge $c\vec{\nabla}\cdot\vec{A}
+\epsilon\partial A^{0}/\partial t=0$ appropriate to the medium.
The $4$-current $J^{\mu}=(\rho/{\epsilon},\vec{J})$, $\rho$.
Since the components of $\vec{J}$ are related by $\vec{J}=\vec{v}\rho$,
we have $\vec{A}=A^{0}{\epsilon}\vec{v}/c$. We calculate
$\vec{A}_{\omega}$ and then use $A^{0}_{\omega}
=\vec{v}\cdot\vec{A}c/(\epsilon v^{2})$. For radiation problems the
denominator factor
$1/|\vec{x}-\vec{x}\,'|$ is replaced by $1/R$. This is standard, with
corrections
of order $a^2/R^2$ or similar effects in the ``near field'' regime,
which is not our subject.

\subsection{Evading The Fraunhofer Approximation}

The Fraunhofer approximation is the textbook expansion for the
phase\\ $\exp{(ik|\vec{x}-\vec{x}\,'|)}
\approx\exp{(ik|\vec{x}|-i k\hat{x}\cdot\vec{x}\,')}$, dropping
terms of order $k|\vec{x}\,'|^{2}/R$. All existing simulations
make this approximation for the phase, and for good reasons: the
subsequent integrations become much simpler. The integrand in
Eq.~(\ref {AA}) oscillates wildly. A Monte Carlo simulation has
to find the surviving phases from myriad cancellations, due to
the phases generated over the length of each track, and then
summed over thousands to millions of tracks moving in three
dimensions. For a $1 TeV$ shower the code of $ZHS$ runs in about
20 minutes on a workstation. Increasing the energy by a factor of
100, the calculational time scales up faster than linear, and computer
time becomes prohibitive. For this reason various strategies to
rescale the output have been used in arriving at the published
values of electric fields. Even for $TeV$ energies, standard Monte
Carlo routines such as GEANT challenge a work-station's capacity.
For cosmic rays of the highest energies the entire approach of
direct numerical evaluation is unfeasible.

Unfortunately the Fraunhofer approximation also neglects terms in the
phases, namely $k|\vec{x}\,'|^{2}/R$, that may be of order unity given
our previous discussion of length and frequency scales.  We must avoid
this step. Progress is possible due to the translational
features of the macroscopic current\\ $J^{\mu}(t',\vec{x}\,')$.
A rather general model is \begin{equation}\vec{J}(t',\vec{x}\,')
=\vec{v}n(z')f(z'-vt',\vec{\rho}\,').\label{J}\end{equation} An even
more general situation will be discussed shortly.  The charge packet
travels with the speed $\vec{v}$, chosen here to be along the $z$-axis
of the coordinate system.  The function $f(z'-vt',\vec{\rho}\,')$
represents a normalized charge density of the traveling packet, with
$\vec{\rho}\,'$ the transverse cylindrical coordinate relative to the
velocity axis.  We normalize $f$ by
$\int\!\!dz'd^{2}\!\rho'\,f(z',\vec{\rho}\,')=1.$ In ice the packet is
about $\Delta z'= 10cm$ thick in the longitudinal direction, and
$\Delta \rho'= 10cm$ in radius (the Moliere radius) in the vicinity of
the shower maximum.  These size scales are limited because of
relativistic propagation.  The time evolution of these scales is
negligible near the shower maximum, and indeed the Moliere radius is
usually approximated by a material constant for the whole shower.
Similarly, in air showers the scale of charge separation is small
compared to the scale of shower longitudinal spread.

The shower's net charge evolution appears in the factor $n(z')$.  With
our normalization, $n(z')$ represents the total charge crossing a
plane at $z'$.  The symbol $n_{ max }$ will denote the maxiumum value
of $n(z')$; later we will see that the electric field scales linearly
with $n_{ max } $.  The longitudinal spread $a$ is a property of
$n(z')$ near the shower maximum.  The model neglects charge (current)
left behind, and moving at less than light-speed in the medium, which
does not emit Cherenkov radiation.  We do not have a sharp cut-off at
the beginning or end of tracks, and the function $n(z')$ will vary
smoothly.

\subsection{Factorization for the Fresnel Zone }

Now while we cannot expand around $\vec{x}\,'=0$ (the Fraunhofer
approximation), the conditions of the problem do permit an
expansion around $\vec{\rho}\,'=0$ (the shower axis), namely for
$R(z')=[(z-z')^{2}+\rho^{2}]^{1/2}$, that

\begin{eqnarray*}|\vec{x}-\vec{x}\,'|&=&[(z-z')^{2}
+(\vec{\rho}-\vec{\rho}\,')^{2}]^{1/2},\\&=&R(z')
-\frac{\vec{\rho}\cdot\vec{\rho}\,'}{R }+{\cal
O}(\frac{\rho'^{2}}{ R}).\end{eqnarray*} For typical values in
this problem, the second term is $\sim10$ times smaller than the
first, and the third is $\sim10^{3}$ smaller than the second. For
the exponent in Eq.~(\ref{AA}), the third term does not
contribute if $k\Delta \rho'^{2}/R\ll 1$, that is $\omega\ll
250GHz$.

Collecting terms, we have \begin{eqnarray}cR\vec{A}_{\omega}= \vec{v}\int
dz'\,n(z')\exp{[i(\frac{\omega}{v}z'+kR)]}\nonumber\\ \times\int\int
dt'\,d^{2}\!\rho'\, \exp{\{-i[\frac{\omega}{v}(z'-vt')+\vec{q}
\cdot\vec{\rho}\,']}\}f(z'-vt',\vec{\rho}\,'). \end{eqnarray} We have
shifted the $t'$ integral which produces the translational phase in the
$z'$ integral. This gives the factorization: 
\begin{equation}\vec{A}_{\omega}\approx
F(\vec{q})\vec{A}^{FF}_{\omega}(\eta),\label{AAA}\end{equation} where
\begin{equation}F(\vec{q})=
\int\!\!d^{3}\!x'\,e^{-i\vec{q}\cdot\vec{x}\,'}f(\vec{x}\,'),
\label{F}\end{equation}
\begin{equation}vcR\vec{A}_{\omega}^{FF}=\vec{v}I^{FF},\end{equation} and
\begin{eqnarray}I^{FF}(\eta, \theta)=\int\!\!dz'\,\exp[\phi(z')],
\nonumber\\ \phi(z')=ik(z'\cos{\theta_{c}}+R(z',\rho) 
)+\log{n(z')}.\label{I}\end{eqnarray} Here $\vec{q}=(\omega/v,
\vec{q}_{\perp})$, $\vec{q}_{\perp}=k\vec{\rho}/R$, and
$\vec{x}\,'=(z',\vec{\rho}\,')$. Provided $F(\omega)\ll 1$ in either
frequency region $k\Delta \rho'^{2}/R\gg 1$ or $k\Delta z'^{2}/R\gg 1$,
the decoupling of the integrals is excellent.  Here $\Delta \rho'$ and
$\Delta z'$ refer to the regions over which the charge exists near the
maximum. $F(\vec{q})$ is the form factor of the charge distribution, which
happens to be defined, just as in the rest of physics, in terms of the
Fourier transform of the snapshot of the distribution. From our
definitions $F(0)=1$.

It is worth noting that the dependence on orientation of
$\vec{q}$ is observable. For example, in a Giant Air Shower,
where the mechanism of charge separation might cause an azimuthal
asymmetry about the shower axis labeled by a dipole $\vec p$,
then $F(\vec{q})$ depends on $F(\vec{q}\cdot\vec{p})$. The
orientation of the dipole relative to the observation point thus
has a strong effect on the emission. (Other numerically large
effects will also be important: for example Allan\cite {Allan}
incorrectly assumes the fields go like $1/R$ from his use of
Feynman's formula).

As a consequence of separating out the form factor, the
integrations have become effectively one-dimensional.

\section{Numerical Work}

At this stage we have a formula for the vector potential which is a
product of a form factor and an object $I^{FF}(\eta, \theta)$
containing the information about the shower history.  We will denote
$I^{FF}(\eta, \theta)$ the {\it Fresnel-Fraunhofer} integral because
it interpolates between these regimes.  In the Fraunhofer
approximation it is easily shown that the factorization is an exact
kinematic feature of translational symmetry as exemplified in
Eq.~(\ref {J}).  If one makes a one-dimensional approximation, the
Fraunhofer integral then evaluates the Fourier transform of the
current\cite {Jackson, newAMZ}.

The factorization in the Fresnel zone is more demanding, yet should be
an excellent approximation.  When calculating $I^{FF}(\eta,\theta)$,
we cannot (as mentioned earlier) consistently expand in powers of
$z'/R$ because $\eta\sim1$ will be needed.

It makes sense at this point to make a numerical comparison with
previous work.  Summarizing the results of an extensive Monte Carlo
calculation in the Fraunhofer approximation, $ZHS$ gave a numerical
fit to the electric field \begin{equation} \frac {R|\vec { E }
^{ZHS}_{\omega}(\theta)|}{F^{ZHS}(\nu)}=1.1\times
10^{-7}\frac{\nu}{\nu_{0}}\frac{E_{0}}{TeV}\exp{[-(\frac{\theta
-\theta_{c}}{\Delta\theta})^{2}]} [\frac{V}{MHz}],\label {zhsfit}
\end{equation} with $\nu_0=500 MHz$.  This is the result of a global
fit to many angles, energies and frequencies $\nu=
\omega/2\pi<\nu_{0}$. In this convention $\omega$ is positive.  The
normalized form factor is $F_{ZHS}(\nu)=1/[1+0.4(\nu/\nu_0)^2]$, as
discussed below.  In making the calculation, results for the field
were also rescaled due to computer limitations.  As a result, the
field reported is strictly linear in the primary energy $E_{0}$.  We
will comment on this shortly.

We calculated our own result proportional to $I^{FF}(\eta=0, \theta)$ over
a range of many frequencies and angles (Figure 2). Before doing the 
integral we scale
out the electromagnetic and dimensional factors which are obvious.  In our
convention $-\infty < \omega < \infty$.  The results of our numerical
integration are quite well fit by
\begin{equation}\frac{R|\vec{E}^{\eta=0}_{\omega}(\theta)|}{F(\omega)}
=\frac{e}{c^{2}}a\sqrt{2\pi}n_{max}\omega
\sin{\theta}\exp{[-\frac{1}{2}(ka)^{2}
(\cos{\theta}-\cos{\theta_{c}})^{2}]}\end{equation} Putting in numerical
values, this gives: \begin{equation}\frac {R|\vec { E }
^{\eta=0}_{\omega}(\theta)|}{F(\omega)}= 2.09\times
10^{-7}\frac{a}{m}\frac{n_{max}}{1000}\frac{\nu}{GHz}
\exp{[-\frac{1}{2}(\frac{\cos{\theta}-\cos{\theta_{c}}}
{\Delta(\cos{\theta})})^{2}]}[\frac{V}{MHz}],\end{equation} where
$$\Delta(\cos{\theta})=0.048\frac{2}{\sqrt{\epsilon}}
\frac{m}{a}\frac{GHz}{\nu}.$$( We have indicated that $a$ is in units
of $m.$ and $\nu$ in $GHz$.) Note the linear dependence on $a \nu$, 
argued earlier to come
from dimensional analysis applied to the limit $\eta \rightarrow 0$.
A cursory inspection shows that this result and the Monte Carlo have
the same general features.

To continue the numerical comparison we need numbers for the
longitudinal spread parameter $a$ and the number of charges at shower
maximum $n_{max}$.  There are several ways to estimate this.  Running
the $ZHS$ code many times and fitting the output of a $1 TeV$ shower
with a cutoff of $611 KeV$ gives $a=1.5m$, $n_{max}=345$. Using these
and allowing for the factor of two in conventions gives agreement to a
few percent in normalization with $ZHS$. However, the other way to do
the calculation is to evaluate the product $an_{max} $ many times.
This method is preferred because fluctuations in $a$ and $n_{max}$ are
correlated. Doing this gives $an_{max}= (570 \pm 50) m$ at $1TeV$, which would
predict a normalization factor of $(1.2\pm 0.1)\times 10^{-7}$ in
Eq.~(\ref{zhsfit}).  This (plus the angular dependence studied below)
indicates that the factorized result is quite consistent with the
Monte Carlo.

In Figure 2 we show numerically integrated values of $I^{FF}$
Eq.~(\ref{I}).  These factors appear directly in the fit just cited,
and serve to check the formulas.  The form factor has been divided
out.  For the range of parameters relevant to the problem, agreement
is very good, and relative error is much less than $1\%$.

For experimental purposes one would like independent confirmation of the
parameters from another source.  Net particle evolution is well described
by Greissen's classic solution, which was simplified further by Rossi to a
Gaussian, $n(z)=n_{max}\,\exp{[-z^{2}/(2a^{2})]}/\sqrt{2\pi}a$.  While
Greissen's $a$ refers to the whole shower, it should also be a reasonable
description for the longitudinal spread of the charge imbalance, which
tends to be a fixed fraction of the total number of particles after a few
radiation lengths.  (There is one caution that the Greissen formula does
not explicitly include low-energy physics important for the charge
imbalance.)  The particular Greissen formula we consulted \cite {Gaisser}
for the longitudinal spread in radiation lengths $X_0$ gives $a/X_{0}\sim
\sqrt { 2/3 log( E_{0}/E ) } $ for particles in the shower with energy
greater than $E$ and a primary with energy $E_{0}$.  In that case one
estimates $a=1.8 m$ at $E_{0}=1 TeV$ with $E =611 MeV, X_{0}= 0.39 m$ in
ice, which is acceptably close to the previous estimates.

At higher energies there is every reason to believe that Greissen's\\
stretched $\sqrt{\log{E}}$ energy dependence will apply.  In that case
$a=2.1m$, $2.3m$ for $100TeV$, $1PeV$ showers, respectively.  Note that
the product $n_{max} a$ is relevant for the field normalization.  In this
case we also need $n_{max}\sim\\1/\sqrt{\log{(E_{0}/E)}-0.33 } $ from the
same Greissen approximation.  Rather amazingly, the product $n_{max}
a\sim(E_{0}/E)\sqrt{2/3\log{(E_{0}/E)}}/\sqrt{\log{( E_{0}/E)}-0.33}\\
\sim E_{0}/E$ at high energies.  This confirms the phenomenon observed by
$ZHS$: the normalization of the electric field (Fraunhofer approximation,
$\theta=\theta_{c}$) scales precisely linearly in the primary energy.  It
is rather pleasing that the result can be understood from first principles
\cite {enrique}.  Later we will see that the parameter $a$ enters in a
much more complicated way in the Fresnel zone, creating an extra, weak
energy dependence. 

Regarding the angular dependence, our work (Figures 3)
indicates a general dependence on $cos{\theta}-cos{\theta_{c}}$
rather than $\theta-\theta_{c}$. When fitting numerical output
the two functional forms are rather different, unless one has a very narrow
distribution. Linearizing for small
$cos{\theta}-cos{\theta_{c}}$ with $a=1.5 m$ for the comparison,
we would predict the scale in the angular dependence
$\Delta\theta=2.1^o(\nu_{0}/\nu) $ while $ZHS$ have the same
expression with $2.4^o$. We find that $\Delta\theta$ is
proportional to $1/a$. If $a$ grows slowly with energy, as
Greissen's formula indicates, then the angular width decreases, which
is not seen in $ZHS$.
Another possible explanation for the small discrepancy is the
improper radiation from tracks terminating abruptly at the ends
used in the Monte Carlo. We have identified these effects as
responsible for the small oscillations seen in the Monte Carlo output,
an effect apparently too small to measure.

When numerical output to the frequency dependence is fit, there is a
slight coupling between the model for the form factor and the
parameter $a$ one will extract for the longitudinal spread.  We made
our own fit to the $ZHS$ code's frequency dependence including the
region up to $1GHz$, using a Gaussian form factor because of its
better analytic properties.  (The $ZHS$ fit, which contains poles in
both the upper and lower half-plane, violates causality.)
Specifically, we find $$F(\nu) =\exp{[-\nu^{2}/(2\nu_{*}^{2})]},\ \ \
\nu_{*}=0.93GHz.$$ Using the corresponding $a$ value we would predict
the Fraunhofer Monte Carlo $\Delta\theta\sim 2.3^o(500
MHz/\nu)$, quite close to $ZHS$.

In real life, shower to shower fluctuations are highly important.  We
studied the statistical features\footnote {We thank Soeb Razzaque for
help with this.} of the parameters $n_{max}$, $a$ by fitting
individual showers many times and looking at the average and $rms$ fit
values.  The results at $E=1TeV$ were $a=1.5 \pm 0.2m$,
$n_{max}=345\pm 60$. Multiplying these and adding fluctuations in
quadratures gives $a n_{max}=520 (1\pm 0.22)m$.  The combination 
$an_{max}*$, which
is the primary variable in determining the normalization of the
electric field, was found to be $(570 \pm 50)m$.  The fluctuation
of $a n_{max}$ is less than half the value that the uncorrelated 
fluctuations of the
separate terms would give.  The relative
fluctuation is said to decrease with increasing energy\cite {enrique},
but there are uncertainties.  For example, threshold rescaling is used
in Monte Carlos, leading to loss of information about the true
fluctuations.  Very preliminary results
of running the standard Monte Carlo GEANT show variations in average
shower parameters such as $a n_{max}$ at the 30$\% $ level compared to
the average of $ZHS$\cite{Soeb}. These comparisons indicate that the
electrodynamics is probably determined better than the rest of the
problem.  Indeed, the deviations from Gaussian behavior in showers is
an effect contributing to the fields at the few percent level.  At the
level of $10-15\%$, many other small effects contribute.  Unless one
uses details about the uncertainties and errors in fits, and
especially about the shower-to shower fluctuations in all relevant
quantities, it is pointless to fine-tune the comparison further.  We
conclude from the numerical work that the factorized expression is at
least as reliable as the Monte Carlos, and has the attractive feature
that the parameters can be adjusted directly.

\section{The Saddle Point Approximation}

While the Rossi-Greissen Gaussian approximation to the shower is
common, there are additional features which favor such an approach to
the emitted radiation.  The coherence is dominated by regions where
the phases add constructively, greatly enhancing the peak region.  In
such circumstances analysis is helpful, especially when the largest
contribution to $I^{FF}$ is dominated by saddle points.  These are
points where the phase is stationary, $d\phi(z_{*})/dz_{*}=0$.

Here we describe the saddle-point method to evaluate $I^{FF}(\eta,
\theta)$.  This is a classic, controlled approximation when the charge
distribution has a single maximum and $kR\gg 1$.  The method turns out to
give the exact result in the limit of flat charge evolution, that is, the
Frank-Tamm formula, where numerical evaluation is highly unstable.  With
the saddle-point approximation, we can extract analytic formulas which are
as good as the numerical integration.

We now describe the saddle-point features.  By translational symmetry the
shower maximum can be located at $z'=0$.  Referring to the formula of
Eq.~(\ref {I}) cited earlier, the saddle point of the phase is given by
solving $$\cos{\theta_{c}}-(z-z_{*})/R(z_{*}) 
-(i/k)(d/dz_{*})\log{n(z_{*})}=0$$ for the point $z_{*}$ dominating the
integral.  The maximum electric field is already known to occur at points
$(z, R)$ near the Cherenkov cone.  For such observation points
$z=R\cos{\theta_{c}}$, and the saddle point equation has an easy solution
at $z_{*}=0$.  Thus the dominant integration region is near the shower
maximum, as physically expected. 

As the point of observation moves off the Cherenkov cone, the saddle
point moves away into the complex $z'$ plane.  To find the complex
saddle-point, we approximate $\log{n(z')}\approx-z'^{2}/(2a^{2})$ in
the vicinity of the shower maximum; that is, we fit {\it the top of
the shower locally} with a Gaussian.  To re-iterate: the saddle-point
approximation does not need to replace the entire shower by a
Gaussian, but replaces the vicinity of the region where phases are
contributing coherently by a Gaussian. The saddle-point condition
gives a quartic equation which can be solved.
Unfortunately the solution is impossibly complicated,
thwarting a direct approach.  We circumvented this by studying the
saddle-point location numerically.  We found the quartic solution is
accurately linearized in a special variable: expand about $\cos{\theta}$
close to $\cos{\theta_{c}}$.  We then find $ z_{*} \approx
R(0)\sin^{2}{\theta} (\cos{\theta}-\cos{\theta_{c}})[1+ i R(0) / ka^{2}
\sin^{2}{\theta} ]^{-1}$.  With this formula the reader inclined can repeat
all the calculations.  We also show the saddle-point to highlight the
appearance of the ratio $ R(0) /( ka^{2}
\sin^{2}{\theta})$ indicated by the qualitative ``acceleration"
argument.  Finally,
replace $R(0)=R=\sqrt{z^{2}+\rho^{2} }$.

The rest of the calculation is standard mathematical physics
\cite{Bender}, so we just quote the results. Calculating fields
from the potential and keeping only the leading terms in
$1/(kR)\ll 1$ we find: \begin{eqnarray}\vec{E}_{\omega}
&=&\frac{i\omega}{Rc^{2}}F(\vec{q})I^{FF}(\eta,\theta)
[(\cos{\theta}-\cos{\theta_{c}})\vec{e}_{R}\nonumber\\
&&-(1-i\eta\frac{\cos{\theta_{c}}}{sin^{2}{\theta}}
\frac{\cos{\theta}-\cos{\theta_{c}}}
{1-i\eta})\sin{\theta}\,\,\vec{e}_{\theta}],
\label{E}\end{eqnarray} \begin{equation}\vec{B}_{\omega}
=-\frac{i\omega}{vcR\cos{\theta_{c}}}F(\vec{q})I^{FF}(\eta,\theta)
(1+i\eta\frac{\cos{\theta}}{sin^{2}{\theta}}
\frac{\cos{\theta}-\cos{\theta_{c}}} {1-i\eta})
\sin{\theta}\,\,\vec{e}_{\phi},\label{B}\end{equation} where
\begin{eqnarray}I^{FF}(\eta,\theta)=\!\!\!\!\!\!&&e^{ikR}a\sqrt{2\pi}
[1-i\eta(\theta)(1-3i\eta\frac{\cos{\theta}}{\sin^{2}{\theta}}
\frac{\cos{\theta}-\cos{\theta_{c}}}{1-i\eta})]
^{-1/2}\nonumber\\&&\exp[-\frac{1}{2}(ka)^{2}
\frac{(\cos{\theta}-\cos{\theta_{c}})^{2}} {1-i\eta}],
\label{II}\end{eqnarray} and where $$\eta=\frac{ka^{2}}{R}
\sin^{2}{\theta}.$$

Inspection reveals that these formulas have the dependence on
$\omega$, $R$, and a distance scale $a$ quoted earlier on physical
grounds. The formulas cited earlier as summarizing the numerical work
are, in fact, the saddle-point approximations evaluated at $\eta=0$,
which fit closer than any empirical formula.

\subsection{Remarks}

The dependence of the fields on symbol $\eta$ summarizes a good deal of
complexity. For example:

\begin{itemize}

\item The limit $\eta \rightarrow 0$ yields the Fraunhofer limit, with
spherical wave fronts and $E_{\omega}\sim a \omega/R.$

\item The limit $\eta \rightarrow \infty$ gives the cylindrically
symmetric $E_{\omega}\sim \sqrt { \omega/R }$ fields. This field can be
substantially different from the Fraunhofer approximation. In fact one
must take this limit to get the Frank-Tamm formula.

A notable application is emission from ultra-high energy air
showers. In such showers the $LPM$ effect plays a definite role
in suppressing the soft emissions from the hardest charges. There
is no corresponding suppression of the evolution of the
low-energy regions of showers where most particles exist, however
\cite {mkrlpm, AZ}. The major effect that we find is that the
showers become long kinematically: that is, the parameter $a$
gets big if one is working with, say, a primary of $10^{20}eV$.
The emitted fields approach those of the Frank-Tamm formula as $a
\rightarrow\infty$, via the formula cited. The effect is
important numerically: a $10^{20}eV$ air shower does not approach the
Fraunhofer limit nearer than $300 km$. The conditions of RICE are
more amenable to the limit, and at $R\sim 1km$, $\omega \sim 1
GHz$ the Fraunhofer-based estimates near the Cherenkov cone in
the literature are good to $20\%$.

\item The frequency dependence in the Fraunhofer approximation is strongly
affected by ``diffraction''. Independent of the form factor effect, the
Fraunhofer approximation imposes an upper limit to the frequency of order
$\omega<(1/a)(\cos{\theta}-\cos{\theta_{c}})$. The true behavior is
substantialli different: from Eq.~(\ref{II}) we see that the field exists
in a region $$\omega<(1/a)[ (\cos{\theta}-\cos{\theta_{c}})^{2}
-(a^{2}/R^{2})\sin^{4}(\theta)]^{-1/2}.$$ For large $\omega$ and in the
angular region where the signal exists, the behaviour is much flatter.
This is illustrated in Figures 4 and 5. This effect is invisible at the
exact value $\theta=\theta_{c}$, where the fake Fraunhofer frequency
cutoff and most of the true functional dependence in the exponent of
$I^{FF}$ both drop out. As a result of the difference in frequency
dependence, the time-structure of the electric field may be substantially
different from the Fraunhofer approximation. We will return to this point
in a Section {\it Causal Features }.

\item Fields in the forward and backward directions,
$\sin^{2}{\theta}\rightarrow 0$, are the fields of $\eta
\rightarrow 0$, the Fraunhofer approximation, regardless of the
physical values of $k, a, R$. We note that the experiments of
Takahashi \cite {Takahashi} observe an extremely limited region
of $\sin^{2}{\theta}\rightarrow 0$. Perhaps this contributes to
the observed agreement with Tamm's formula in a regime where $a^2 k/R$ is
not close to zero.

\item The polarization varies considerably.  From symmetry the
polarization is in the plane of the charge and the observation point.
Moreover, for $\theta=\theta_{c}$ the electric field is transverse to
$\vec{R}$ for any $\eta$.  Yet naive transversality is not true in general
at any finite $\eta$.

\item Between the various limits the dependence on every scale in the
problem, namely the frequency $\omega$, the distance $R$, the length scale
$a$, and the angle $\theta$, is neither that of the Fraunhofer limit nor
that of the infinite track Frank-Tamm limit, but instead a smooth
interpolation between the cylindrical and spherical wave regimes.

\item  In the finite $\eta$ limit, one may also include a further
effect, namely that as $\omega\rightarrow 0$ one has a `near-zone'
Coulomb-like response at small $R$. (Indeed, the
$\omega\rightarrow0$ limit measures the net charge.) This effect,
important below about $10MHz$, also has a slight effect on the time
structure of pulses.

\end{itemize}

We pause to comment on the generality of the result.  What if we had
not made the physical, but specific ansatz (\ref{J})?  The entire
analysis can be repeated for an arbitrary charge distribution
$j(t',z',\vec{\rho}\,')$.  The Fraunhofer expansion of the transverse
variable, and the Fourier integral of the $t'$ variable are general.
Provided the $\vec{\rho}\,'$ extent is finite, and there exists a
dominant $z'$ region, then the integrals always factor into a
product of a form factor and a one-dimensional integral for
$A^{FF}_{\omega}(\eta)$.  In fact nothing changes (the reader can
repeat the calculation) except that when an arbitrary current is
set up, the existence of a single saddle point cannot be assumed.
Corrections to the local Gaussian approximation are
straightforward.  The slight skewness of real showers (or other
arbitrary charge distributions) can also be developed as a
saddle-point power series.  Again: there are elements of
bremmstrahlung in real showers, having a stochastic nature, which
the current model has not attempted to reproduce.  Detailed Monte
Carlo simulations \cite{Soeb} of our group have included the
bend-by-bend amplitudes of tracks undergoing collisions.  This
goes well beyond the approximation of a single straight line
track, suddenly beginning and ending, of the previous literature
\cite{ZHS,AZ}.  The effect of all the small kinks is negligible
except in the very high frequency region $\omega\gg 100GHz$,
while the endpoint accelerations give oscillations in the angular
dependence down by orders
of magnitude.  As a final side remark: we explicitly studied
contributions of finite tracks, just to see what would happen, in the
development towards the conditions of the Tamm formula.
It is straightforward to develop these pieces if one needs them
for, say, the Takahashi-type experiments \cite {Takahashi}, in
the Fresnel zone.

\subsection{The General Case}

We now turn to fields valid for any $\eta$. The form factor,
which was extracted from the Fraunhofer calculations, is
universal and need not be changed. The validity of the
saddle-point approximation does not depend explicitly on the
value of $\eta$. The procedure of linearization to locate the
saddle-point happens to be good to
$cos{\theta}-cos{\theta_{c}}\sim 1$, so that the approximation is
rather good in the entire region $R/a\gg 1$, $kR\gg 1$.

For practical applications it is useful to have a formula for the fields
with quantities measured in physically motivated units. For this purpose
we rewrite (\ref{E}) as follows
\begin{equation} R\vec{E}_{\omega}\approx
2.52\cdot10^{-7}\frac{a}{m}\frac{n_{max}}{1000}\frac{\nu}{GHz}
F(\vec{q})\,\psi\,
\vec{\cal{E}}\,\,[\frac{V}{MHz}].\label{EE}\end{equation} Here $n_{max}$ is the
excess of electrons over positrons at the shower maximum, $R$ is measured
in meters, $\nu$ is measured in $GHz$.  The
rescaled field is \begin{eqnarray}
\vec{\cal{E}}&=&[-\frac{\cos{\theta}-\cos{\theta_{c}}}{\sin{\theta}}
\vec{e}_{R}+(1-i\eta\frac{\cos{\theta_{c}}}{sin^{2}{\theta}}
\frac{\cos{\theta}-\cos{\theta_{c}}}{1-i\eta})
\,\,\vec{e}_{\theta}]\nonumber\\
&&[1-i\eta(1-3i\eta\frac{\cos{\theta}}{\sin^{2}{\theta}}
\frac{\cos{\theta}-\cos{\theta_{c}}}{1-i\eta})]^{-1/2}\nonumber\\
&&\exp[-\frac{1}{2}(ka)^{2}\frac{(\cos{\theta}-\cos{\theta_{c}})^{2}}
{1-i\eta}].\label{epsilon}\end{eqnarray} We have defined a kinematic
factor $\psi =-i\exp(ikR)\sin{\theta}$ in such a way that the rescaled
field $\vec{\cal{E}}$ is normalized at $\theta=\theta_{c}$,
\begin{equation}\vec{\cal{E}}(\theta=\theta_{c})=(1-i\eta)^{-1/2}
\,\,\vec{e}_{\theta}.\end{equation}

It is convenient to plot the magnitude of the rescaled field,
Eq.~(\ref{epsilon}).  Figure 3 shows the magnitude
as a function of the angle difference $\theta-\theta_{c}$ in various
limits. The Fraunhofer approximation is shown by a dashed curve, and
our result by the solid curve. One observes that the Fraunhofer limit
is approached from below. This is physically clear: The Fresnel zone
fields have a wider angular spread, and conservation of energy forces
them to be smaller in magnitude compared to the sharper,
diffraction-limited Fraunhofer fields. As the fields evolve to
infinity, they coalese into narrower and taller beams.

The frequency dependence of $R|\vec{E}_{\omega}|/F(\omega)$ is
shown on Figures 4 and 5. Exactly at the Cherenkov angle the
difference between the Fraunhofer approximation and our results are
minor for the typical parameters of $RICE$.  However, away from the
Cherenkov angle there is a substantial difference between the two,
throughout the region where the magnitude of the field is large.  This
effect can be masked by the form factor, so we have plotted $R
E_{\omega}/F(\omega)$ to show it.  This effect may have important
repurcussions for the time-structure of pulses, which are also
discussed in the last Section.

Figures 6-10 are contour plots of the electric field. We did not bother
to remove the small region $a/R\sim 1$, where our result does not apply. The
Fraunhofer approximation has trivial $1/R$ dependence on the distance to
the observation point (Figure 10). The exact result is certainly 
different, with
Figures 6, 8, and 9, in particular, illlustrating the effects of
constructive interference
in the region of cylindrical symmetry. A complementary view examines
contour plots of constant phase. This is shown in Figures 8 and 9. In
making these figures we decreased the kinematic phase
$ikR$ to values showing several oscillations (as opposed to hundreds)
across the range of
the plots. The lines
of constant phase can be used to illustrate the time-evolution of waves of
a given frequency: that is, the Fourier transform of $\delta
(\omega-\omega_{*}) E_{\omega}exp(-i\omega t)$ has wave fronts at each
moment in time given by the lines of constant phase. The constant phase
lines of the Fraunhofer approximation are, of course, spherical
(Figure 10). The
constant phase lines of the true behaviour interpolate between cylindrical
and spherical symmetry. As a consequence of the Fresnel-zone behaviour,
the eikonals of the expanding radiation field do not emerge radially,
but actually curve due to interference effects. This is a sobering impact
of very basic physics, which has a measurable effect in the signal
propagation speed discussed later under the topic of causal feature.

\section{Causal Features}

With our convention that the electric field $E(t,\vec{x})
=1/(2\pi)\int_{-\infty}^{\infty}\!\!d\omega\,\exp(-i\omega 
t)\\E_{\omega}(\vec{x})$,
causality requires $E_{\omega}$ to be analytic in the upper
half-plane of $\omega $. Singularities in the lower half-plane
determine the details of $E(t,\vec{x})$ and the precise causal
structure.

To discuss this we consider detection of signals via an antenna-system
response function $\cal A_{\omega}$.  By standard arguments the
detected voltage is a convolution in time, and therefore a product in
$\omega$ space, of the antenna function and the perturbing electric
field.  The antenna function has the same causal analytic properties
as the electric field.  A proto-typical antenna or circuit function
for a driven $LRC$ circuit is $$ {\cal
A}_{\omega}=Z /(-\omega^{2} +\omega_{\cal A}^{2}-i\Gamma
\omega), $$ where $Z $ depends on where the amplifier is
connected in the circuit and can be treated as a constant.  Note that 
$\cal A_{\omega}=
-Z /(\omega-\omega_{+})(\omega-\omega_{-})$ where
$\omega_{\pm}$ are in the lower half-plane.  The dielectric function
$\epsilon(\omega)$ can then be taken as slowly varying in the region
where the antenna and form factor allow a response, and also has its
analytic structure in the lower half-plane if this detail needs to be
included. We will also ignore the form factor for this discussion,
which earlier was
cited as a formula analytic in the complex plane.  While nothing in
our analysis depends on these
idealizations, this approach to the analytic
structure serves to make our point.

As a first illustration, consider the electric field fit given by
$ZHS$, proportional to $1/(\omega-i\omega_{0})(\omega+i\omega_{0})$
with $\omega_{0}/(2\pi)=500 MHz$.  This field has poles in both the
upper and lower half-plane, violating causality.  One may argue that
the literal analytic behavior in the complex plane goes beyond the
ambitions of the original semi-empirical fit.  Nevertheless, Cauchy's
theorem applies to the subsequent numerical integrations that have
been made \cite {fmr}, giving non-causal branches to
numerically evaluated Fourier transforms, as well as unphysical
short-time structure.

Let us compare the analytic structure of the electric field in
the saddle-point approximation. This approximation
does not attempt to describe the region $\omega \rightarrow 0$, which
requires treatment of the near zone. However for causality we do not
need $E_{\omega}$ near the origin but at large $|\omega|$. The
saddle point approximation is good here so the results should be
reliable.

Let us investigate this in more detail. In the exponent in the
expression for the electric field we have $-(1/2)(ka)^{2}
(\cos{\theta}-\cos{\theta_{c}})^2(1+i\eta)/(1+\eta^{2})$. Since $\eta$
goes like $\omega$ there is a phase linear in $\omega$ at large
$|\omega|$. There is
also a branch-cut and pole from the prefactor which occurs at
$\eta=-i$. All singularities are in the lower half-plane
and consistent with causality. The causal structure of
$E(t,\vec{x})$ then hinges on closing the contour at infinity. For
this the details of the antenna function, which generally
has isolated singularities, as well as singular behavior of
$E_{\omega}$ near the origin
do not matter.

We close the contour at infinity avoiding the branch cuts, which may be
oriented along the negative imaginary axis or as consistent. Convergence
then requires $$ \lim_{\omega\rightarrow-i\infty}\mathrm{Re}[-i\omega
t+i\omega\sqrt{\epsilon}R-\frac{1}{2}\omega^{2}\epsilon a^{2}(
\cos{\theta}-\cos{\theta_{c}})^{2}\frac{i\eta}{1+\eta^{2}}]<0.$$ Using
the
definition of $\eta$, causality implies $$ t-\sqrt{\epsilon}\frac{R}{c}
[1-\frac{(\cos{\theta}-\cos{\theta_c})^{2}}
{2\sin^2{\theta}}]^{-1}>0.$$

This result has a natural interpretation.  While the radiation from the
shower appears to come primarily from the geometric location of the
maximum, $\theta\sim \theta_{c}$, the shower actually develops and
radiates earlier.  Consequently the strict causal limit must correspond to
an apparent propagation speed slightly faster than the naive speed of
$c/\sqrt{\epsilon}$ deduced from the location of the observation point at
$R$.  The earliest signal actually arrives at an apparent speed $v_{app}$
of $$v_{app}=\frac{c}{\sqrt{\epsilon}}[1-\frac
{(cos{\theta}-cos{\theta_{c}})^{2}}{2\sin^{2}{\theta}}]^{-1}.$$ This
formula is entirely geometrical, consistent with the simple picture that
distance differences in the problem are causing the effect, but it also
incorporates subtle features of coherence.  For example, the distance
scale $a$ cancels out.  Yet $a$ determines the angular spread
$\cos{\theta}-\cos{\theta_{c}}$ over which most of the power in the wave
is contained, and enters in this fashion.

The Fraunhofer limit, in which all signals originate at a single point of
origin, is incapable of capturing such an effect.  It is interesting to
trace the origin of the discrepancy.  The singularities of interest are
located by the zeroes of $1-i\eta=1-i\omega\sqrt{\epsilon }\sin^{2}
{\theta}/R.$ When the limit $R \rightarrow \infty$ is taken in the first
step of the Fraunhofer approximation, all the non-trivial analytic
structure moves away to $\omega \rightarrow -i\infty$ and is lost.  This
procedure does not commute with closing the contour at $|\omega
|\rightarrow \infty$.  The correct procedure, of course, is to first close
the contour, and then take the limit of large $R$. 

In practice, of course, not all of the signal arrives at the earliest
possible moment.  The time scale over which the signal is detected depends
on competition between dispersion, the antenna and form factor details,
and something like twice the "advanced" time interval.  This time interval
is $\Delta t_{caus}=R\sqrt{\epsilon}/(2c)
(\cos{\theta}-\cos{\theta_{c}})^{2}/\sin^{2}{\theta}$.  Since the $\Delta
t_{caus } $ effect scales proportional to the distance $R$, it does not
become negligible in any limit, exhibiting another subtle facet of
breakdown of the Fraunhofer approximation. 

\vspace{1cm} {\bf Acknowledgments:} This work was supported in part by
the Department of Energy, the University of Kansas General Research
Fund, the\\ K*STAR programs and the \textit{Kansas Institute for
Theoretical and Computational Science}.  We thank Jaime
Alvarez-Mu\~{n}iz, Enrique Zas, Doug McKay, Soeb Razzaque and Suruj
Seunarine for many helpful suggestions and conversations.  We
especially thank Enrique for writing and sharing the ZHS code, and
Soeb and Suruj for generously sharing their results from running shower
codes.

\begin{center}
\begin{figure}
\epsfxsize=16cm
\epsfbox{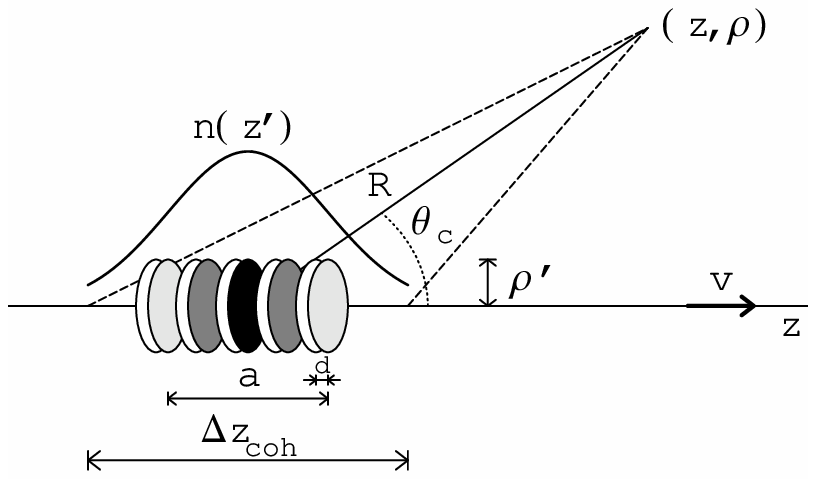}

\caption {Cartoon of electromagnetic coherence zone associated with an
evolving shower. The shower develops over a longitudinal spread dimension
$a$, as represented by the smooth curve.  During most of this development
the pancake of charge has a constant size $d$, but an evolving charge
normalization, represented by the gray scale. The coherence zone $\Delta
z_{ coh }$ depends on the angle, frequency, and location of the
observation point. The situation illustrated has $\Delta z_{ coh }>a$, a
case in which full coherence of the charge (up to a frequency cutoff
determined by the pancake size)  is obtained.  In the reverse limit
$\Delta z_{coh}<a$, only the portion of the charge inside the region of
$\Delta z_{coh}$ contributes constructively. }

\end{figure}
\end{center}

\begin{center}
\begin{figure}
\epsfxsize=15cm
\epsfbox{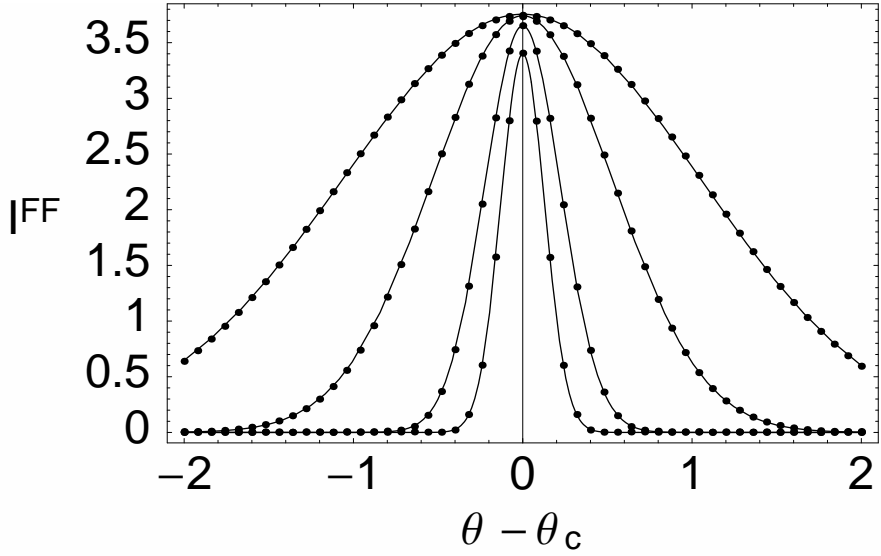}

\caption {Comparison of numerically integrated (points) and analytic fit
(solid curve) to the Fresnel-Fraunhofer integral $I^{FF}(\eta,\theta)$,
Eq.~(\ref{I}). Values of parameters are: $a=1.5m$, $R=1000m$; frequency
$\nu=1,2,5,10GHz$ (from top to bottom). Both Fresnel and Fraunhofer
regimes are successfully reproduced.  Angle is measured in degrees,
$I^{FF}$ in meters.}

\end{figure} 
\end{center}

\begin{center}
\begin{figure}
\epsfxsize=16cm 
\epsfbox{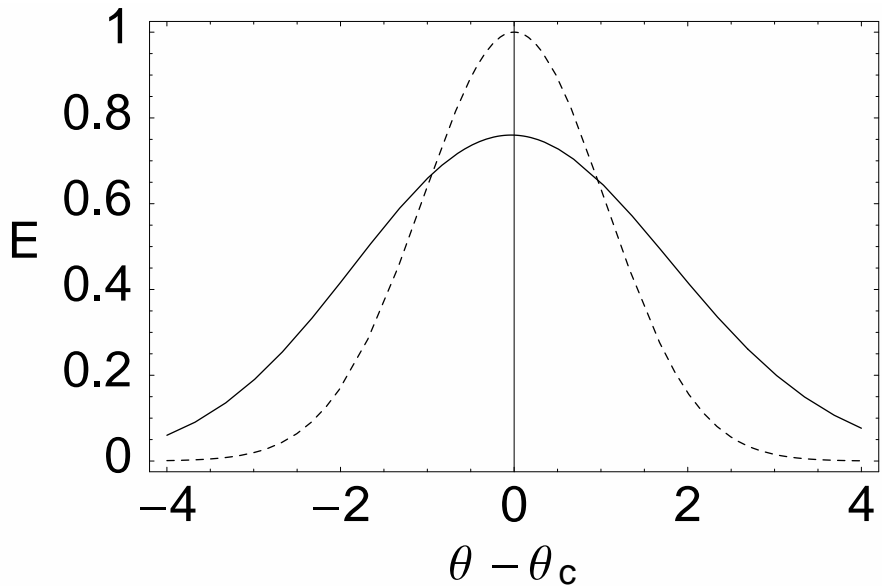} 

\caption{Magnitude of the rescaled electric field $E=|\vec{{\cal E}}|$,
Eq.~(\ref{epsilon}).  Our result (solid curve) is compared to the
Fraunhofer approximation (dashed curve).  $a=1.5m$, $R=50m$, $\nu=1GHz$;
angle is measured in degrees.  The Fraunhofer approximation is narrower in
angular extent and larger in magnitude.  Due to conservation of energy the
field sharpens its angular distribution and approaches the asymptotically
far field from below as $R\rightarrow \infty$.  Up to scaling
normalization factors, the same plot would apply to any with the same
$\eta$, for example $a=3m$, $R=200m$.}

\end{figure}
\end{center}

\begin{center} 
\begin{figure} 
\epsfxsize=15cm 
\epsfbox{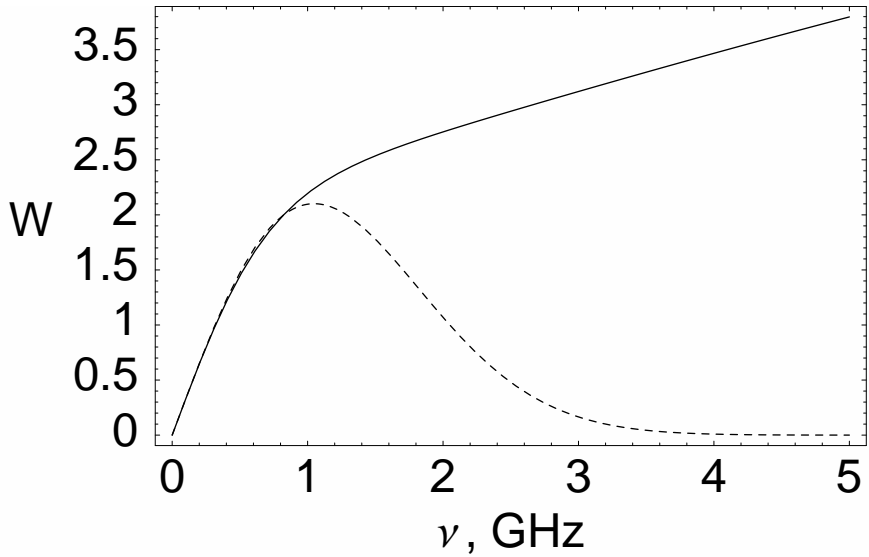}

\caption{Magnitude of the re-scaled electric field $W=
(1000/n_{max})R|\vec{E}_{\omega}|/F(\omega)$ as a function of frequency
$\nu$, evaluated at $\theta=\theta_{c}+1^{o}$. $W$ is measured in
$10^{-7}V/MHz$. The dashed curve is the Fraunhofer approximation. Note the
high frequency cutoff, which is not a form factor effect, but an
artificial result of the limit $R\rightarrow\infty$ imposed by the
approximation. Solid curve is our result. The parameters $a=1.5m$,
$R=100m$.}

\end{figure} 
\end{center}

\begin{center}
\begin{figure}
\epsfxsize=15cm
\epsfbox{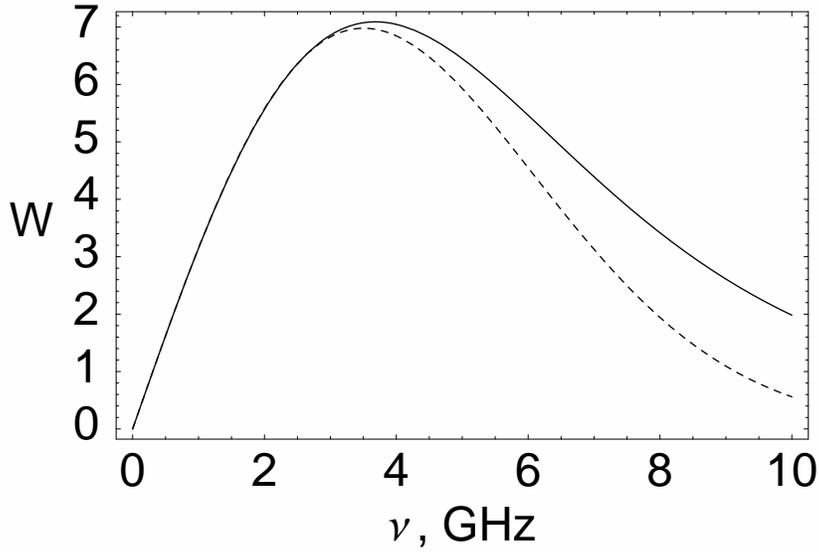}

\caption{Magnitude of the re-scaled electric field $W=
(1000/n_{max})R|\vec{E}_{\omega}|/F(\omega)$ as a function of frequency
$\nu$, evaluated at $\theta=\theta_{c}+0.3^{o}$. $W$ is measured in
$10^{-7}V/MHz$. The dashed curve is the Fraunhofer approximation; solid
curve our result. As $R\rightarrow\infty $ the Fraunhofer approximation
begins to apply. The parameters $a=1.5m$, $R=1000m$.}

\end{figure}
\end{center}

\begin{center}
\begin{figure}
\epsfxsize=15cm
\epsfbox{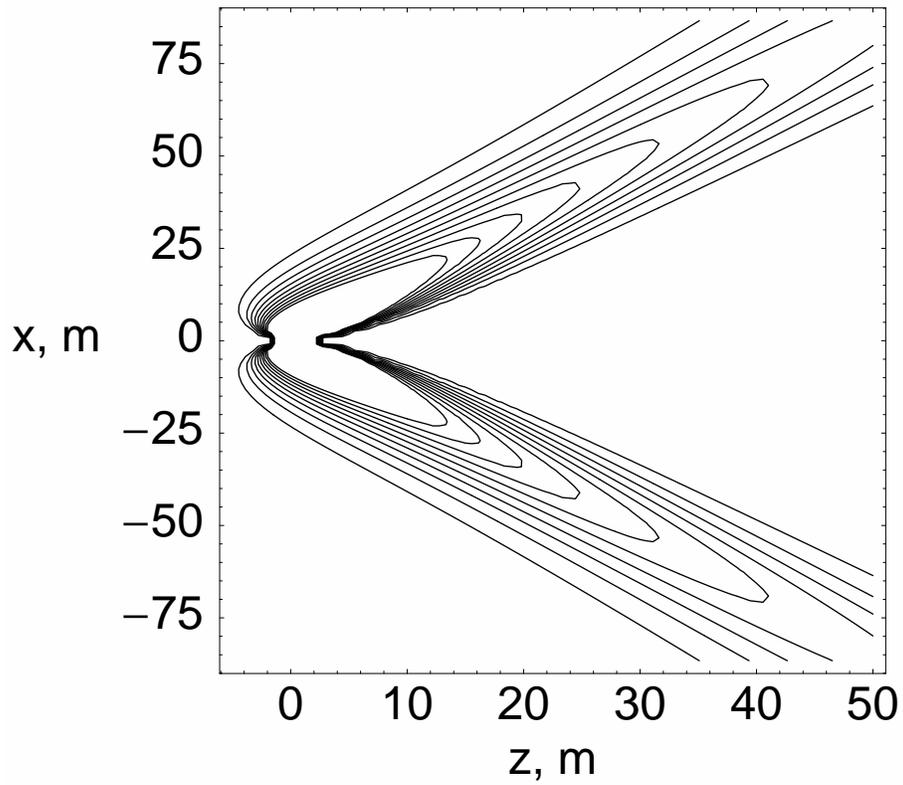}

\caption{Contour plot for the magnitude of the electric field. $a=5m$,
$\nu=100MHz$, distances in $m$. Note the evolution of the field from
cylindrical to spherical behaviour as the distance from the origin
increases.}

\end{figure}
\end{center}

\begin{center}
\begin{figure}
\epsfxsize=15cm
\epsfbox{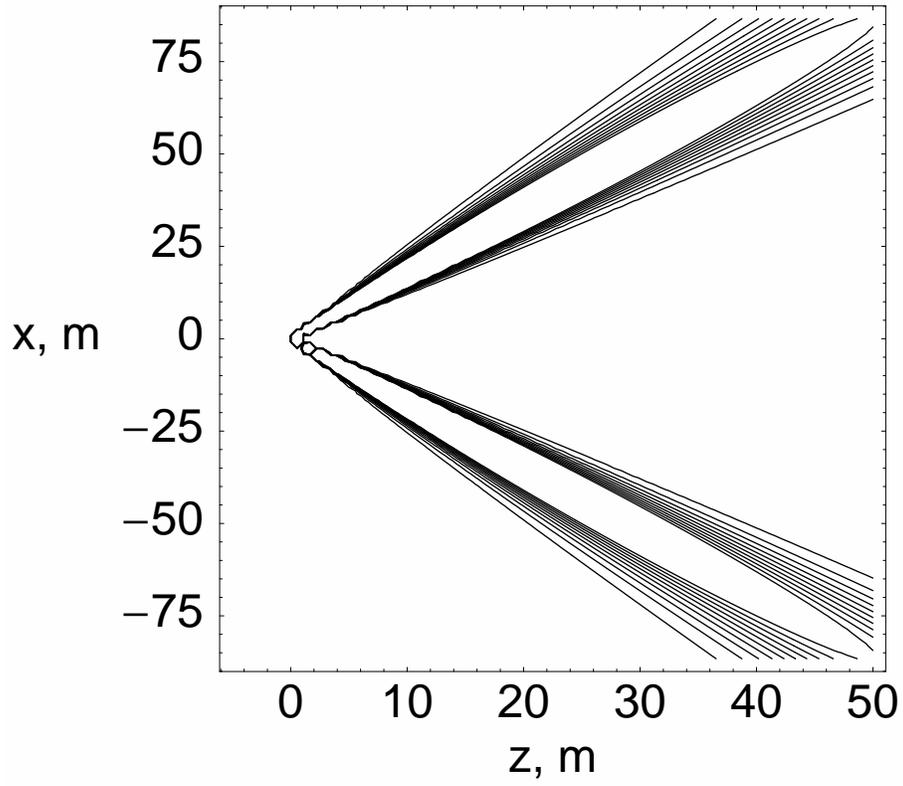}

\caption{Contour plot for the magnitude of the electric field in the
Fraunhofer approximation. The parameters $a=5m$, $\nu=100MHz$; distances
in $m$. The magnitude of the field lacks the richness of structure of
Figure 6.}

\end{figure}
\end{center}

\begin{center}
\begin{figure}
\epsfxsize=15cm
\epsfbox{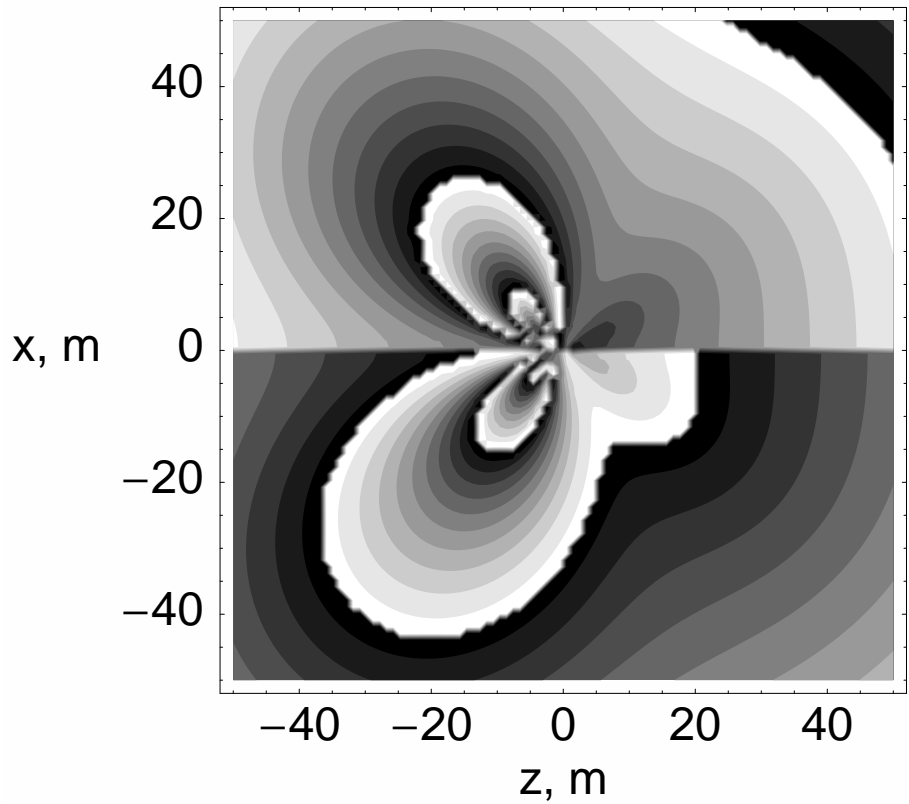}

\caption{Contour plot for the phase of the electric field. The parameters
$a=1.5m$, $\nu=100MHz$; distances in $m$. Each contour represents the time
evolution of an expanding wave front for the frequency used. The phases
illustrate non-trivial propagation with destructive interference, as well
as evolution from cylindrical to spherical symmetry.}

\end{figure}
\end{center}

\begin{center}
\begin{figure}
\epsfxsize=15cm
\epsfbox{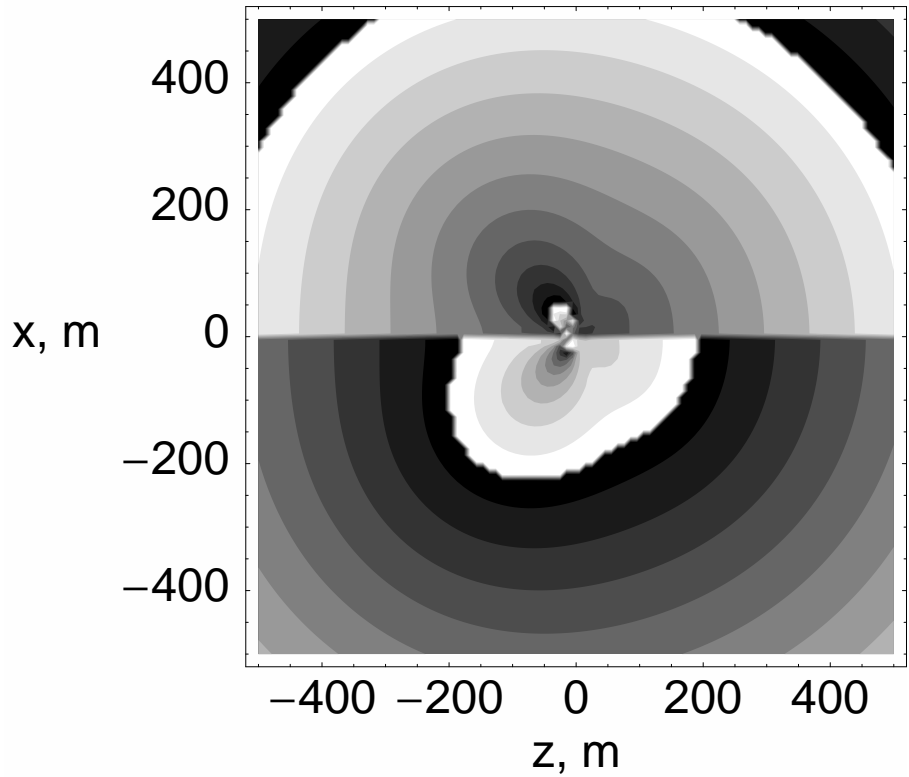}

\caption{Contour plot for the phase of the electric field. The parameters
$a=1.5m$, $\nu=100MHz$;  distances in $m$. The phases illustrate
non-trivial propagation with destructive interference, here shown on a
larger scale than Figure 8.}

\end{figure}
\end{center}

\begin{center}
\begin{figure}
\epsfxsize=15cm
\epsfbox{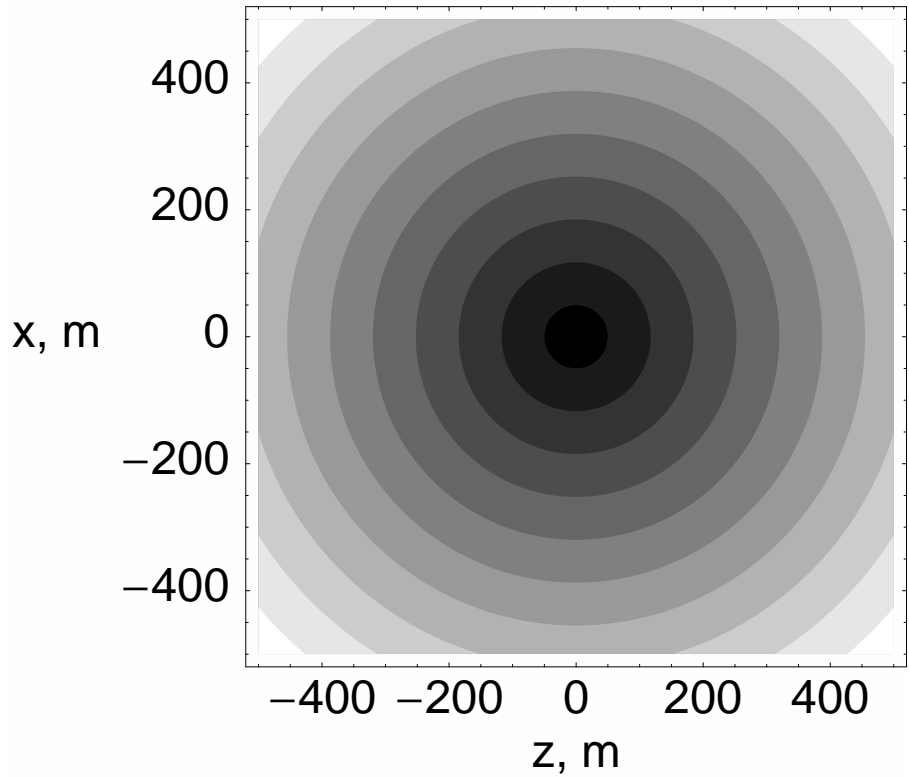}

\caption{Contour plot for the phase of the electric field in the
Fraunhofer approximation with $a=1.5m$, $\nu=100MHz$; distances in $m$.
The contours illustrate kinematic spherical symmetry imposed by the
approximation. }

\end{figure}
\end{center}

\end{document}